\documentclass[11pt]{article}
\usepackage{times}
\usepackage{geometry}
\usepackage[utf8]{inputenc}
\usepackage{enumitem,amssymb}
\usepackage{ragged2e}
\usepackage{graphicx}
\usepackage{comment}
\usepackage{multicol}
\usepackage{wrapfig}
\usepackage[usenames]{xcolor} 
\definecolor{xlinkcolor}{cmyk}{1,1,0,0}
\usepackage{url}
\usepackage[
 colorlinks=true,    
 linkcolor=xlinkcolor,     
 citecolor=xlinkcolor,     
 filecolor=xlinkcolor,  
 urlcolor=xlinkcolor,      
 final=true
]{hyperref}
\usepackage[super,sort&compress]{natbib}
\usepackage{enumitem}
\setenumerate{itemsep=0mm}

\begin{document}
\begin{raggedright} 
\huge
Snowmass2021 - Letter of Interest \hfill \\[+1em]
\textit{Radio Detection of Ultra-high Energy Cosmic Rays with Low Lunar Orbiting SmallSats} \hfill \\[+1em]
\end{raggedright}

\normalsize

\noindent {\large \bf Thematic Areas:}  (check all that apply $\square$/$\blacksquare$)

\noindent $\blacksquare$ (CF1) Dark Matter: Particle Like \\
\noindent $\blacksquare$ (CF6) Dark Energy and Cosmic Acceleration: Complementarity of Probes and New Facilities \\
\noindent $\blacksquare$ (CF7) Cosmic Probes of Fundamental Physics \\

\noindent {\large \bf Contact Information:}\\
Andr\'es Romero-Wolf (Jet Propulsion Laboratory, California Institute of Technology) [Andrew.Romero-Wolf@jpl.caltech.edu]: \\
Collaboration (optional): Zettavolt Askaryan Polarimeter (ZAP) \\

\noindent {\large \bf Authors:} 
Andr\'es Romero-Wolf$^a$,
Jaime Alvarez-Mu\~niz$^b$, 
Luis A. Anchordoqui$^c$, 
Douglas Bergman$^d$,
Washington Carvalho Jr.$^e$,
Austin L. Cummings$^f$,
Peter Gorham$^g$,
Casey J. Handmer$^a$,
Nate Harvey$^a$,
John Krizmanic$^h$, 
Kurtis Nishimura$^g$,
Remy Prechelt$^g$,
Mary Hall Reno$^i$, 
Harm Schoorlemmer$^j$,
Gary Varner$^g$,
Tonia Venters$^k$, 
Stephanie Wissel$^l$, 
Enrique Zas$^b$
\\[+1em]
$^a$Jet Propulsion Laboratory, California Institute of Technology, 
$^b$IGFAE \& Universidade Santiago de Compostela, 
$^c$Lehman College, City University of New York, 
$^d$University of Utah,
$^e$Universidade do S\~ao Paulo
$^f$Gran Sasso Science Institute,
$^g$University of Hawai'i at M\=anoa, 
$^h$University of Maryland, 
$^i$University of Iowa, 
$^j$Max Planck Institute, 
$^k$NASA Goddard Space Flight Center, 
$^l$Pennsylvania State University, 
\\[+1em]
\noindent {\large \bf Abstract:} 
Ultra-high energy cosmic rays (UHECRs) are the most energetic particles observed and serve as a probe of the extreme universe. A key question to understanding the violent processes responsible for their acceleration is identifying which classes of astrophysical objects (active galactic nuclei or starburst galaxies, for example) correlate to their arrival directions. While source clustering is limited by deflections in the Galactic magnetic field, at the highest energies the scattering angles are sufficiently low to retain correlation with source catalogues. While there have been several studies attempting to identify source catalogue correlations with data from the Pierre Auger Observatory and the Telescope Array, the significance above an isotropic background has not yet reached the threshold for discovery. It has been known for several decades that a full-sky UHECR observatory would provide a substantial increase in sensitivity to the anisotropic component of UHECRs. There have been several concepts developed in that time targeting the identification of UHECR sources such as OWL, JEM-EUSO, and POEMMA, using fluorescence detection in the Earth's atmosphere from orbit. In this white paper, we present a concept called the Zettavolt Askaryan Polarimeter (ZAP), designed to identify the source of UHECRs using radio detection of the Askaryan radio emissions produced by UHECRs interacting in the Moon's regolith from low lunar orbit.  

\bigskip

\noindent{\it Acknowledgment: The research was carried out at the Jet Propulsion Laboratory, California Institute of Technology, under a contract with the National Aeronautics and Space Administration (80NM0018D0004). \copyright 2020. All rights reserved. Pre-Decisional Information – For Planning and Discussion Purposes Only.}

\clearpage


\noindent{\Large \bf Introduction -}
The physical origin of ultra-high energy cosmic rays (UHECRs) beyond the well-established spectral cutoff energy of $\sim 10^{19.6}$~eV~\cite{Abbasi:2007sv,Abraham:2008ru,Abraham:2010mj,Aab:2020rhr,Aab:2020gxe}
is currently a topic of debate. 
While spatial clustering of events with astrophysical sources is an obvious way of identifying the sources, this is complicated by the fact that Galactic magnetic fields scatter the cosmic ray arrival directions~\cite{Farrar:2017lhm}. 
%
%
At energies beyond the cutoff the scattering scale is $\sim 15^\circ$, which does allow to test whether classes of sources are correlated to anisotropies in the distribution of UHECR arrival directions.

\begin{figure}[!b]
    \centering
    \includegraphics[width=1.\textwidth]{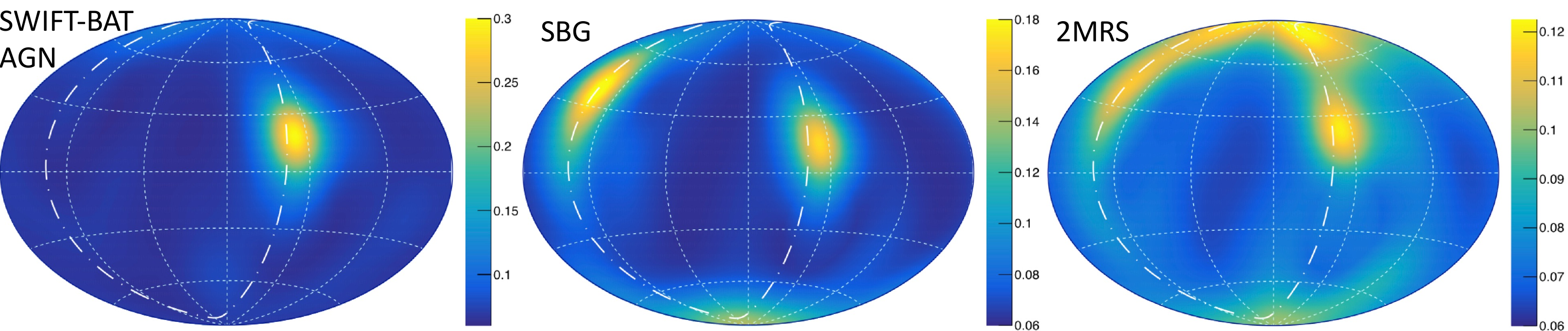}
    \caption{Source distributions assuming a scattering angle of 15$^\circ$, as expected from Galactic magnetic fields, with an anisotropic fraction of UHECR events of 20\%. From left to right the source catalogues are SWIFT-BAT Active Galactic Nuclei, Starburst Galaxies, and the 2MRS~\cite{Aab:2018chp}. Note that the color scales in each map (in Galactic coordinates) have different ranges. }
    \label{fig:skymaps}
\end{figure}

There are several compelling classes of source candidates~\cite{Kotera:2011cp,Anchordoqui:2018qom,Sarazin:2019fjz}
and as can be seen in Figure~\ref{fig:skymaps}, some of these have discernible sky 
distributions~\cite{Aab:2018chp}. 
One hypothesis is that the particle acceleration are due to shocks in the backflows of active galactic nuclei (AGN) jets, which are powered by the accretion of supermassive black holes~\cite{Biermann:1987ep,Rachen:1992pg,Matthews:2018rpe}. 
A second candidate is the termination shocks in starburst galaxies (SBGs), driven by intense star formation rate~\cite{Anchordoqui:1999cu,Anchordoqui:2018vji,Anchordoqui:2020otc}. 
Some scenarios for UHECR acceleration that result in sky distributions that correlate to the matter distribution in the nearby universe are the unipolar induction in rapidly rotating magnetospheres of newly born pulsars~\cite{Fang:2012rx} 
 and the shocks in gamma-ray bursts~\cite{Waxman:1995vg,Vietri:1995hs,Globus:2014fka,Zhang:2017moz}.

Both the Pierre Auger Observatory (Auger), in the southern hemisphere and the Telescope Array (TA) in the northern hemisphere have been reporting excesses of events over expectations from an isotropic sky in angular regions spanning about $20^\circ$. TA has recorded an excess above the isotropic background-only expectation in cosmic rays with energies above $10^{19.75}~{\rm eV}$~\cite{Abbasi:2014lda,Abbasi:2020fxl}. Auger has performed a correlation search with various catalogues (SBGs, $\gamma-$AGN, Swift-BAT, and 2MRS) and has found that the largest significance over anisotropy is $4.5\sigma$ for SBGs~\cite{Aab:2018chp,Aab:2019ogu}. TA 
 has reported that with their current statistics they cannot make a statistically significant corroboration or refutation of the possible correlation between UHECRs and SBGs~\cite{Abbasi:2018tqo}.

To overcome the challenges imposed by using two different observatories with uneven sky coverage and different systematic effects to identify the sources of UHECRs, it is necessary to have a single observatory with full-sky nearly-uniform coverage. This has been the motivation behind several concepts for space-based observation of UHECRs using fluorescence detectors in low-Earth orbit (OWL~\cite{Stecker:2004wt}, JEM-EUSO~\cite{Adams:2013vea}, and most recently POEMMA~\cite{Anchordoqui:2019omw}). While the performance of these concepts has several attractive qualities, they are resource-intensive and their implementation is challenging. 

An alternative implementation for a full-sky coverage UHECR detector is to use radio detectors in low lunar orbit. In the 1960's Askaryan proposed that cosmic rays entering the surface of the Moon and generating particle showers would generate strong coherent radio emission~\cite{Askaryan:1962hbi,Askaryan:1965hbi}. This effect has been verified in several accelerator experiments in a number of dense dielectric media and has been shown to match microscopic simulations 
~\cite{Saltzberg:2000bk,Gorham:2004ny}.
Askaryan radiation is the basis for several current~\cite{Allison:2015eky,Anker:2019rzo,Allison:2019xtn}
and proposed ~\cite{Alvarez-Muniz:2018bhp,Wissel:2020sec}
observatories including NASA's ANITA mission~\cite{Gorham:2019guw}.
The Lunar Orbiting Radio Detector (LORD) mission concept~\cite{Ryabov:2016fac} has been previously proposed to search for ultra-high energy neutrinos and has shown to have high sensitivity to UHECRs. In this white paper, we present the Zettavolt Askaryan Polarimeter (ZAP), which uses radio detection from low lunar orbit with a focus on determining the sources of ultra-high energy cosmic rays. In addition, ZAP offers unprecedented sensitivity to UHECRs with energies $> 10^{20}$~eV, which can be used to discriminate between energy cutoff hypotheses. Designing a lunar radio detector for UHECRs results in additional benefits including sensitivity to UHE neutrinos and photons that may be produced by the decay of super-heavy dark matter particles. 

\bigskip

\noindent{\Large \bf The Zettavolt Askaryan Polarimeter (ZAP) - }
\begin{wrapfigure}{R}{0.55\linewidth}
    \centering
    \includegraphics[width=0.55\textwidth]{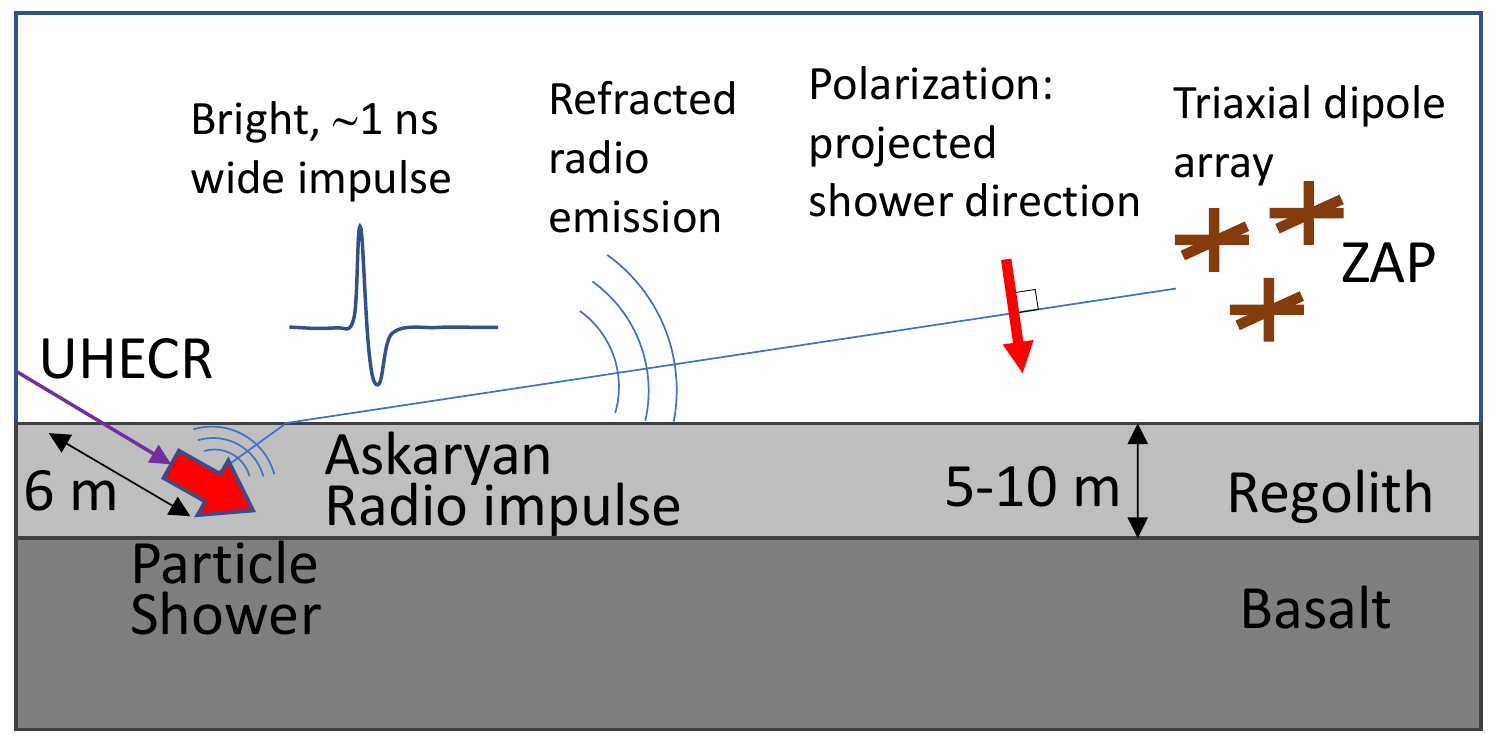}
    \caption{\label{fig:zap1} ZAP measures the energy and direction of arrival from UHECRs incident on the lunar surface with detection of the electric field strength, spectrum, and polarization (see text for details).  }
    \label{fig:my_label}
\end{wrapfigure}
ZAP is designed with the objective of determining which classes of objects accelerate UHECRs by correlating the anisotropic component to candidate source catalogues (SBG, AGN, or 2MRS); see Figure~\ref{fig:skymaps}. The basic observatory concept is shown in Figure~\ref{fig:zap1}. An ultra-high energy cosmic ray enters the lunar regolith to produce an air shower that will reach shower maximum within the first $\sim 6$~m. The Askaryan radio emission is an impulsive transient with a beam pattern that varies from dipole-like at low frequencies ($\sim 30$~MHz) and transitions to Cherenkov cone-like at high frequencies ($\sim300$~MHz), providing an observation angle-dependent radio spectrum. The signal is 100\% linearly polarized with a polarization vector that is determined by the direction of the UHECR shower projected in a plane orthogonal to the line of sight. The signal refracts out of the Lunar surface to be detected by an array of wideband active dipoles ($30-300$~MHz). The direction of the signal is reconstructed by the combination of differences in the timing of arrival between spatially separated dipoles, the polarization vector, and the pulse frequency spectrum. The CR energy is determined from the amplitude of the electric field and the reconstructed position and direction of the UHECR-induced shower in the lunar regolith.  

An implementation of ZAP in low lunar orbit ($\sim 100$~km altitude) can achieve an event rate of $> 1000$ for energies $>10^{19.6}$~eV in two years of operation. The details of the mission implementation, to optimize for the pointing and energy resolution needed to achieve statistically significant correlation to source catalogues, are currently under study. With ZAP's sensitivity to UHECRs above $10^{20}$~eV, it is also possible to distinguish between models where the UHECR spectrum ends due to the “end-of steam” of cosmic accelerators~\cite{Allard:2008gj} or due to UHECR interaction with the cosmic microwave background~\cite{Greisen:1966jv,Zatsepin:1966jv}. In the latter a recovery of the proton spectrum is expected~\cite{Anchordoqui:2019omw}. This imparts a test of Lorentz Invariance with unprecedented precision~\cite{Stecker:2017gdy}. In addition, ZAP can explore a space of models for superheavy dark matter of UHE neutrinos and photons~\cite{Berezinsky:1997hy}. UHE neutrinos are significantly more deeply penetrating than UHECRs and would manifest themselves as upward-going showers from beneath the regolith. UHE photon have a comparable but larger  penetration depth than UHECRs but they would manifest themselves as having a train of multiple radio pulses due the LPM effect~\cite{AlvarezMuniz:1997sh}.



\bigskip

\noindent{\Large \bf Conclusions - }
The ZAP mission concept is potentially a low-resource implementation to determine the sources of ultra-high energy cosmic rays. With air shower arrays the size of small nations that have already been operating for more than a decade, the next major step in UHECR observatories are space-based implementations with large exposures achieved in relatively short mission durations. These observatories would not also expand the reach of UHECRs to energies well above the cutoff, but they also have the potential to put the most stringent limits on super heavy dark matter by way of detecting ultra-high energy neutrinos and photons. ZAP would target a launch within this decade.

\clearpage

\bibliographystyle{utphys}
\bibliography{snowmass_ZAP}

\vspace{3.5in}


\end{document}